\documentclass[runningheads]{llncs}
 \pdfoutput=1

\usepackage{graphicx}
\usepackage{subcaption}
\usepackage{calc}
\usepackage[normalem]{ulem}
\usepackage{amsmath}
\usepackage{amssymb}
\usepackage{algorithm}
\usepackage[noend]{algpseudocode}
\algnewcommand\algorithmicinput{\textbf{Input:}}
\algnewcommand\INPUT{\item[\algorithmicinput]}
\algnewcommand\algorithmicoutput{\textbf{Output:}}
\algnewcommand\OUTPUT{\item[\algorithmicoutput]}
\newcommand{\norm}[1]{\lVert#1\rVert}
\usepackage{pdfpages}
\usepackage{tabularx,booktabs}
\newcolumntype{Y}{>{\centering\arraybackslash}X}
\usepackage{float}
\usepackage[utf8]{inputenc}
\usepackage{pgfplots}
\usepackage[misc]{ifsym}

\begin{document}
	
\title{Optimal Edge User Allocation in Edge Computing with Variable Sized Vector Bin Packing\thanks{The final authenticated publication is available online at https://doi.org/10.1007/978-3-030-03596-9\_15}}
\titlerunning{Optimal Edge User Allocation in Edge Computing}

\author{Phu Lai\inst{1} \and Qiang He\inst{1}$^{\textrm{(\Letter)}}$ \and Mohamed Abdelrazek\inst{2} \and Feifei Chen\inst{2} \and John Hosking\inst{4} \and John Grundy\inst{3} \and Yun Yang\inst{1}}

\authorrunning{P. Lai et al.}
\institute{
	Swinburne University of Technology, Hawthorn, Australia \\
	\email{\{tlai,qhe,yyang\}@swin.edu.au} \and
	Deakin University, Burwood, Australia \\
	\email{\{mohamed.abdelrazek,feifei.chen\}@deakin.edu.au} \and
	Monash University, Clayton, Australia \\
	\email{john.grundy@monash.edu} \and
	The University of Auckland, Auckland, New Zealand \\
	\email{j.hosking@auckland.ac.nz}
}

\maketitle

\begin{abstract}
	In mobile edge computing, edge servers are geographically distributed around base stations placed near end-users to provide highly accessible and efficient computing capacities and services. In the mobile edge computing environment, a service provider can deploy its service on hired edge servers to reduce end-to-end service delays experienced by its end-users allocated to those edge servers. An optimal deployment must maximize the number of allocated end-users and minimize the number of hired edge servers while ensuring the required quality of service for end-users. In this paper, we model the edge user allocation (EUA) problem as a bin packing problem, and introduce a novel, optimal approach to solving the EUA problem based on the Lexicographic Goal Programming technique. We have conducted three series of experiments to evaluate the proposed approach against two representative baseline approaches. Experimental results show that our approach significantly outperforms the other two approaches.
	
\end{abstract}

\section{Introduction}
In recent years, the world has witnessed a surge in the number of cloud and mobile network connected end-devices, including mobile phones, wearables, sensors and a wide range of Internet of Things (IoT) devices. According to Ericsson's Mobility Report \cite{Ericsson2017}, it is predicted that there will be around 32 billion of such connected devices by 2023. This has produced a great challenge for online service providers in terms of guaranteeing a reliable and low-latency connection to end-users, which is one of the key quality-of-service (QoS) requirements \cite{Varghese2017}.

To tackle this issue, Cisco \cite{Bonomi2012} has proposed the fog computing paradigm -- also called \emph{edge computing} -- in which computation, storage, and networking resources are pushed closer to the edge of the network by deploying a number of intermediate \emph{edge servers} with closer proximity to \emph{end-devices}. This paradigm offers lower network latency and greater scalability than the conventional centralized cloud computing paradigm. This is particularly important for high volume streaming applications or critical systems such as autonomous traffic systems, health care, or cloud gaming, which require real-time decision making. In edge computing, online service providers hire existing edge servers to host their services to serve their end-users. Thin clients -- such as wearables, sensors or smart phones -- all that have limited storage and computing capability, benefit from this architecture by the capability to offload intensive computing tasks to the distributed edge servers near them \cite{Yi2015}. In this way, the central cloud is not required to perform all the computing tasks single-handedly, which is highly resource demanding and generally incurs long network latency for end-users. Usually, an edge server covers a specific geographical area so that the users within its coverage can connect to it via LTE, 4G or Radio Network \cite{Hu2015}. A number of edge servers would be deployed in a distributed fashion (usually near cellular base stations \cite{Hu2015}) so that they can cover different geographical areas. The coverages of adjacent edge servers usually partially overlap to avoid blank areas not covered by any edge server.  A user located in the overlapping area can connect to one of the edge servers covering them (\emph{proximity constraint}) that has sufficient computing resource (\emph{capacity constraint}) such as CPU, bandwidth, or memory. 

Edge servers' capacity, current workloads, coverages, the number of users to allocate and their proximity to end-users can be obtained or calculated at any time. Based on this information, while fulfilling the above constraints, an optimization goal must be achieved from a service provider's perspective -- to minimize the number of edge servers used -- in order to attain an optimal solution to the allocation of the service provider's users due to the pay-as-you-go pricing model applied in edge computing \cite{Yi2015,Varghese2017}, which might incur higher costs when the number of edge servers used increases. Additionally, due to the aforementioned constraints, there might be a number of users that cannot be assigned to any edge servers. Those users will be connected directly to a central cloud server. Therefore, another optimization objective is to maximize the number of users allocated to hired edge servers.

We refer to the above problem as an \emph{edge user allocation} (EUA) problem then model it as a \emph{variable sized vector bin packing} (VSVBP) problem, a non-geometric generalization of the classical {\sc bin packing } (BP) problem. The EUA problem is critical in edge computing, however, has not been properly investigated. Solutions to the task allocation problem in cloud computing have been investigated in \cite{Ren2017,Wolke2015}. However, the edge computing architecture is different from cloud computing, i.e., distributed vs. centralized. In addition, the various constraints and dynamic information discussed above significantly differentiate the edge computing environment from the traditional cloud computing environment with many unique characteristics. Thus, the approaches for task allocation in cloud computing are not suitable for solving the EUA problem, hence the need for a new approach. In this paper, we make the following major contributions:
\begin{itemize}
	\item we have modeled and formulated the EUA problem as a VSVBP problem;
	\item we have developed an optimal approach for solving the EUA problem using the Lexicographic Goal Programming technique; and
	\item we have evaluated our approach against two representative baseline approaches with extensive experiments to demonstrate its effectiveness.
\end{itemize}

The remainder of the paper is organized as follows. Section 2 motivates this research with an example. In section 3, we give a background of the VSVBP problem. Section 4 discusses the proposed approach, which is evaluated in section 5. Section 6 reviews the related work. Section 7 concludes this paper.

\section{Motivating Example}

A representative example of edge computing applications is large-scale mobile gaming \cite{Lin2017a} - the fastest growing gaming model \cite{BI2017}. The cloud gaming model has made online game platforms, such as Hatch\footnote{https://www.hatch.live/} and Sony PlayStation Now\footnote{https://www.playstation.com/en-gb/explore/playstation-now/}, more accessible for thin-client mobile players since the resource-expensive game instance is running on a remote cloud server. Consider an increasingly popular virtual reality game G, which requires a great amount of computing power for graphic rendering. Employing the traditional centralized cloud model helps thin-clients offload the heavy computation tasks; however, this approach introduces a huge network delay due to the long distance between players and cloud servers. Therefore, pushing the processing power closer to players with edge computing is a promising solution to this problem. Fig. 1 shows an example of edge computing architecture that can be implemented in this scenario.

\setlength\belowcaptionskip{-2ex}
\begin{figure}
	\centering
	\includegraphics[page=1,scale=0.7]{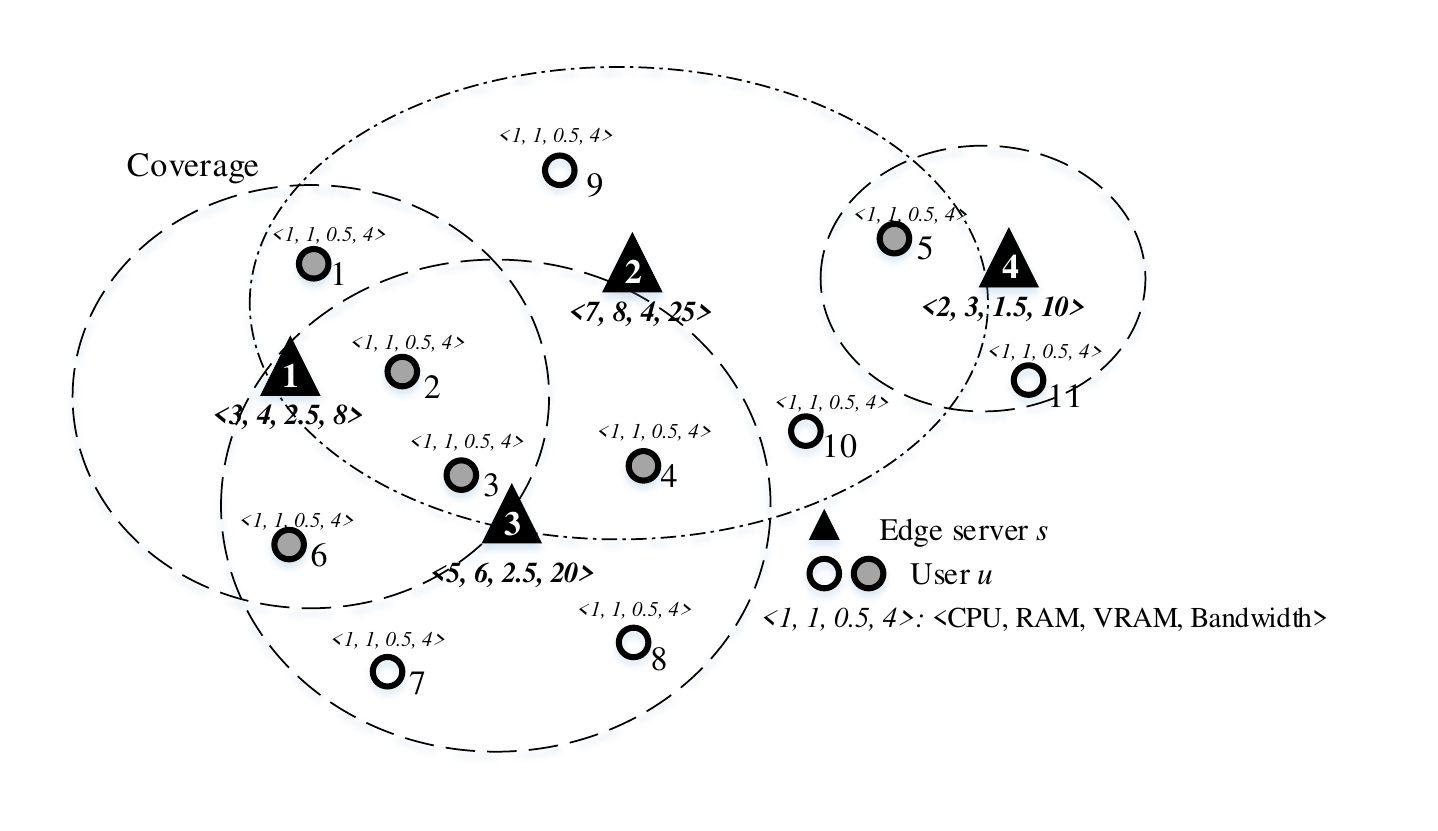}
	\caption{Edge computing deployment example}
\end{figure}

Assume there are four edge servers in a specific area that can be used to host game G. Each edge server covers a particular geographical area. Users who are outside the coverage of an edge server will not be able to connect to it (\textit{proximity constraint}). For example, user $ u_{4} $ cannot be assigned to edge server $ s_{1} $ or $ s_{4} $ and has to be allocated to either server $ s_{2} $ or $ s_{3} $. Furthermore, we need to take into account various \emph{capacity constraints} such as bandwidth, memory, processing power, etc. In Fig. 1, each edge server has a limited computing capacity denoted as a vector $ \langle CPU core, memory, VRAM, bandwidth \rangle $. The aggregate workload generated by users on a server must not exceed the remaining capacity of that server. There are seven users within the coverage of edge server $ s_{2} $ with a total workload of $ \langle 7, 7, 3.5, 28 \rangle $, exceeding the remaining capacity of server $ s_{2} $ ($ \langle 7, 8, 4, 25 \rangle $). Thus, the game provider cannot assign all of these users to a single server $ s_{2} $. Since users $ u_{1} $, $ u_{2} $,...,$ u_{5} $ are also covered by other edge servers, it is possible to allocate them to other servers to share the workload with server $ s_{2} $. One potential solution would be to allocate users $ u_{1} $, $ u_{2} $ to server $ s_{1} $, users $ u_{3} $, $ u_{6} $ to server $ s_{3} $ and users $ u_{4} $, $ u_{5} $ stay with server $ s_{2} $. No proximity or resource constraint is violated this way, but this might not be the optimal solution. If we assign users $ u_{1} $, $ u_{2} $, $ u_{4} $ to server $ s_{2} $,  users $ u_{3} $, $ u_{6} $ to server $ s_{3} $, and user $u _{5} $ to server $ s_{4} $, server $s _{1} $ will no longer be required so the service provider can choose not to hire it to lower the total cost of hiring edge servers. This solution satisfies all the aforementioned constraints, uses the least servers to serve the most users, as well as guarantees the QoS.
\section{Background}
\textbf{Definition 1. Classical Bin Packing (BP) Problem.} \textit{Given an infinite supply of identical bins $ S = \{s_{1},s_{2},...,s_{i}\} $ with maximum capacity $ C = 1 $ and a set of $ n $ items $ U = \{ u_1, u_2, ..., u_j \} $. Let a value $ w_j \equiv w(u_j) $ be the size of item $ u_j $ that satisfies $ 0 < w_j \leq C $ and $ 1 \leq j \leq n $. The objective is to pack all the given items into the fewest bins possible such that the total item size in each bin must not exceed the bin capacity $ C $: $ \textstyle \sum_{u_{j} \in U(s_{i})} w_{j} \leq C, \forall s_{i} \in S $.}

In the classical BP problem, one can normalize $ C = 1 $ without loss of generality since the bin capacity is just a scale factor. Aggregating item sizes not exceeding the capacity of the corresponding bin is the only constraint. This problem is known to be an $\mathcal{NP}$-hard combinatorial optimization problem \cite{Garey1979}.

\textbf{Definition 2. Variable Sized Bin Packing (VSBP) Problem.} \textit{Given a limited collection of bin sizes such that $ 1 = size(s_1) > size(s_2) > ... > size(s_k) $, there is an infinite supply of bins for each bin type $ s_k $. Let $ L = \{ s^1, s^2,...,s^l \} $ be the list of bins needed for packing all items. Given a list of items $ U = \{ u_1, u_2, ..., u_j \} $ with $ size(u_j) \in (0,1] $, the objective of the VSBP problem is to find an item-bin assignment so that the total size of the bins required $ \sum_{b=1}^{l}size(s^b) $ is minimum.}

In the classical BP problem, all bins are homogeneous with a similar bin capacity. VSBP is a more general variant of the classical BP in which a limited collection of bin sizes is allowed. VSBP aims at minimizing the total size of the bins used, which is slightly different compared to the objective of the classical BP problem as discussed above.

\textbf{Definition 3. Vector Bin Packing (VBP) Problem.} \textit{Given a set of $ n $ items $ U = \{ u_1, u_2, ..., u_j \} $, the size of an item $ u_j $ is denoted as a d-dimensional vector $ w_{j} = \langle w_{j}^1, w_{j}^2, ..., w_{j}^d  \rangle $, $ w_j \in [0,1]^d $. One is given an infinite supply of identical bins $ S = \{s_{1},s_{2},...,s_{i}\} $  with maximum capacity $ C=\langle 1^1, 1^2,...,1^d \rangle $. The objective is to pack the set $ U $ into a minimum number of bin $ s $ such that $ \norm{\sum_{u_{j} \in U(s_{i})} w_{j}}_\infty \leq 1, \forall s_{i} \in S $.} 

In the classical BP problem, the size of an item is presented as a single aggregation measure. By contrast, the size of an item in the VBP problem is associated with a multi-dimensional vector. The objective remains similar, in which the sum of packed item size vectors must not exceed the bin capacity vector in each dimension, which is normalized to 1 without loss of generality. The VBP problem is also known as multi-capacity BP problem in some literature. 

In the EUA problem, a bin is referred to as an edge server with the maximum bin capacity being the remaining computing resource of that edge server. An item is referred to as an edge user, which can be a mobile phone or any IoT device; the size of an item is the amount of workloads generated by that user, measured by the computing resource needed to perform the requested task. In this paper, we tackle the EUA problem from a service provider's perspective. Thus, all users of an application generate the same amount of workload. In the real world, different edge servers may have different hardware specifications and host different applications for different numbers of users. Thus, they have different remaining server capacities, or computing resources. In addition, a computing task has various resource requirements such as CPU core, memory, video RAM, bandwidth, and so forth. Therefore, the amount of computing resource needed for a task should not be calculated by a just a single aggregate measure; instead, it can be represented as a $ d- $dimensional vector where each dimension represents a resource type. The proposed VSVBP problem for EUA is $\mathcal{NP}$-hard since the classical BP, which is $\mathcal{NP}$-hard \cite{Garey1979}, is a special case of VSVBP where $ d = 1 $ and all the bins are identical in their capacity dimensionality.
\section{Our Approach}

\subsection{Definitions}

Edge servers have differentiated remaining capacity and multi-dimensional resource requirements for computation tasks. Therefore, the EUA problem can be modeled as a mixture of the VSBP problem and the VBP problem, hence a variable sized vector bin packing (VSVBP) problem. Our objective is to maximize the number of allocated users and minimize the number of hired edge servers.

We first introduce relevant notations and definitions used in our model in Table 1. In the EUA problem, every user covered by any edge server must be allocated to an edge server unless all the servers accessible to the user have reached their maximum capacities. If a user cannot be allocated to any edge servers, or is not positioned within the coverage of any edge servers, they will be directly connected to a service provider's central cloud server.

\begin{table}[H]
	\caption{Key Notations}
	\label{table_notations}
	\begin{tabular}{|l|p{9cm}|}
		\hline
		\textbf{Notation} & \textbf{Description} \\
		\hline
		$ S = \{s_{1},s_{2},...,s_{i}\} $ & finite set of edge server $ s_{i} $, where $ i = {1,2,...,m} $ \\
		\hline
		$ C_{i} = \langle C_{i}^1, C_{i}^2, ..., C_{i}^d  \rangle$ & $ d- $dimensional vector with each dimension $ C_{i}^k $ being a resource type, such as CPU utilization or disk I/O, representing the remaining capacity of an edge server $ s_{i} $, $ k \in \{1,2,...,d\} $ \\
		\hline
		$ U = \{u_{1},u_{2},...,u_{j}\} $ & finite set of user $ u_{j} $, where $ j = {1,2,...,n} $ \\
		\hline
		$ w_{j} = \langle w_{j}^1, w_{j}^2, ..., w_{j}^d  \rangle $ & $ d- $dimensional vector representing the size of the workload incurred by user $ u_{j} $. Each vector component $ w_{j}^k $ is a resource type, $ k \in \{1,2,...,d\} $ \\
		\hline
		$ U(s_{i}) $ & set of users allocated to server $ s_{i} $. $ U(s_{i}) \subset U $ \\
		\hline
		$ d_{ij} $ & geographical distance between server $ s_{i} $ and user $ u_{j} $ \\
		\hline
		$ cov(s_{i}) $ & coverage radius of server $ s_{i} $ \\
		\hline
	\end{tabular}
\end{table}

The total workload generated by all users allocated to an edge server must not exceed its remaining capacity (1). Otherwise, the server will be overloaded, causing service disruptions or performance degradation. Take Fig. 1 for instance. The aggregate workload incurred by users $ u_{5} $ and  $ u_{11} $ is $ \langle 2, 2, 1, 8 \rangle $ does not exceed the remaining capacity of server $ s_{4} $, $ \langle 2, 3, 1.5, 10 \rangle $; therefore, this is a valid user-server assignment. If we allocate users $ u_{1}, u_{2}, u_{3}, u_{4}, u_{5}, u_{9}, u_{10} $ to server $ s_{2} $, it will be overloaded since the aggregate user workload $ \langle 7, 7, 3.5, 28 \rangle $ exceed the server's remaining capacity $ \langle 7, 8, 4, 25 \rangle $.
\begin{displaymath}
\sum\limits_{u_{j} \in U(s_{i})} w_{j} \leq C_{i}, \quad \forall s_{i} \in S \tag*{(1)}
\end{displaymath}
In the classical BP problem, an item can be placed in any bins as long as the bin has sufficient remaining capacity. However, in our problem, an edge server covers a limited surrounding geographical region. Thus, an item (user) can be assigned to specific bins (edge servers) since an edge server can only serve users who are located within its coverage (2). Take Fig. 1 for example. Server $ s_4 $ can serve users $ u_5$ and $ u_{11} $ only. Since users might position in the overlapping areas of different edge servers, there is an optimal solution to allocate as many users as possible to as few servers as possible, which is the main focus of our research.
\begin{displaymath}
d_{ij} \leq cov(s_{i}), \hfill \forall i \in \{ 1,2,...,m \}; \forall j \in \{ 1,2,...,n \} \tag*{(2)}
\end{displaymath}
Our primary objective is to maximize the number of users allocated to all hired edge servers, which ensures the QoS from the service provider's perspective:
\begin{displaymath}
maximize \hspace{5pt} \sum\limits_{s_i \in S} |U(s_i)|, \hfill  \tag*{(3)}
\end{displaymath} 
Our secondary objective is to find a user-server assignment $ \{ u_1,...,u_j \} \longrightarrow \{ s_1,...,s_i \} $ such that the number of servers hired $ E $  is minimum:
\begin{displaymath}
minimize \hspace{5pt} E =  |\{ s_i \in S | \sum\limits_{u_{j} \in U(s_{i})} w_{j} > 0 \}| \tag*{(4)}
\end{displaymath}

\subsection{EUA Model}

In this paper, we address the EUA problem with two optimization objectives: 1) maximizing the number of users allocated and 2) minimizing the number of edge servers hired, while satisfying the \textit{capacity constraint} and \textit{proximity constraint}. Accordingly, we model the EUA problem as a Lexicographic Goal Programming (LGP) problem \cite{romero2014}. In a lexicographic goal program, there are multiple optimization objectives with a number of constraints. These objectives are ranked by their levels of importance, or priorities. The solver will attempt to find an optimal solution that satisfies the primary objective and then proceed to find a solution for the next objective without deteriorating the previous objective(s). An LGP program can be solved as a series of connected integer linear programs. The LGP formulation of the EUA problem is as follows:

\begin{flalign}
\indent \indent \indent &maximize \ \sum\limits_{j=1}^{n}\sum\limits_{i=1}^{m} x_{ij} \tag{5} &\\ 
\indent \indent \indent &minimize \ E = \sum\limits_{i=1}^{m} y_{i} \tag{6}
\end{flalign}
subject to:
\begin{align}
&\sum\limits_{j=1}^{n} w_{j}^{k}x_{ij} \leq C_{i}^{k}y_{i}, \forall i \in \{ 1,...,m \}; \forall k \in \{1,...,d\}  \tag{7} &\\ 
&d_{ij} \leq cov(s_{i}), \forall i \in \{ 1,...,m \}; \forall j \in \{ 1,...,n \} \tag{8} &\\ 
&\sum\limits_{i=1}^{m} x_{ij} \leq 1, \forall j \in \{ 1,...,n \} \tag{9} &\\ 
&y_{i} \in \{ 0, 1 \}, \forall i \in \{ 1,...,m \} \tag{10} &\\ 
&x_{ij} \in \{ 0, 1 \}, \forall i \in \{ 1,...,m \}; \forall j \in \{ 1,...,n \} \tag{11}
\end{align}
where: 

\indent \indent \indent $ y_{i} = 1 $ if server $ s_{i} $ is hired. \\
\indent \indent \indent $ x_{ij} = 1 $ if user $ u_{j} $ is allocated to server $ s_{i} $. \\
\indent \indent \indent $ cov(s_{i}) $ is provided by edge computing providers. \\

The objective (5) maximizes the number of users that are assigned to hired edge servers. The objective (6) minimizes the number of hired edge servers. Here, objective (5) is ranked higher than objective (6) in terms of priority. There are two groups of binary variables, i.e., $ x_{ij} $ (11) and $ y_i $ (10).

\textit{Capacity constraint}: As described by (7), each edge server $ s_{j} $ has a remaining capacity of $ C_{i} = \langle C_i^1, C_i^2,...,C_i^d, \rangle $, a d-dimensional vector. The aggregate workload of each resource type incurred by all allocated users must not exceed the corresponding remaining capacity of their assigned server. Take Fig. 1 for example. Assigning users $ u_5, u_{11} $ to server $ s_4 $ is valid since $ \langle 2, 2, 1, 8 \rangle < \langle 2, 3, 1.5, 10 \rangle $.

\textit{Proximity constraint}: As described by (8), only users located within the coverage of an edge server can be allocated to the edge server. A user may be located in the overlapping coverage of multiple edge servers. For instance, users $ u_2, u_3 $ can be allocated to servers $ s_1, s_2$ or $ s_3 $.

Constraint family (9) ensures every user is allocated to at most one edge server. In other words, a user can be allocated to either an edge server or service provider's cloud server.

\section{Experimental Evaluation}
In this section, we evaluate the performance of our approach by extensive experiments with a comparison to two baseline approaches. All the experiments were conducted on a Windows machine equipped with Intel Core i5-7400T processor (4 CPUs, 2.4GHz) and 8GB RAM. The LGP problem modeled in section 4.2 was solved using IBM ILOG CPLEX Optimizer.

\subsection{Baseline approaches}
Our approach will be benchmarked against two baseline approaches for user-to-server assignment, namely \emph{random} and \emph{greedy} approaches:
\begin{itemize}
	\item \emph{Random}: Each user will be allocated to a random edge server as long as that server has sufficient remaining capacity to accommodate the user and has the user within its coverage.
	\item \emph{Greedy}: Each user will be allocated to an edge server that has the most remaining capacity and has the user within its coverage.
\end{itemize}

\subsection{Experiment settings}
In this paper, we conduct experiments on data of base stations and end-users within the  Melbourne central business district area in Australia, which has a total area of 6.2 km\textsuperscript{2}. 

\textbf{Experiment data:} We collect the location data of edge servers and end-users. Australian Communications and Media Authority (ACMA) publishes the radio-comms license dataset that contains the geographical location of all cellular base stations in Australia, which we will use as the locations of edge servers \cite{Hu2015}. The coverage of each edge server is randomly set within a range of 450 - 750 meters. In terms of end-users' locations, Asia Pacific Network Information Centre (APNIC) provides all IP address blocks allocated to Australia. We use an IP lookup service\footnote{http://ip-api.com/} to convert the obtained IP addresses into geographical locations. Since IP addresses in the last octet are likely to have identical geographical addresses returned by the IP lookup service, more end-users are uniformly generated around each of the obtained geographical locations. The raw experimental data has been made publicly available (EUA-dataset\footnote{https://github.com/swinedge/eua-dataset}).

\textbf{Experimenting parameters:} In the experiments, we vary three setting parameters that may have an impact on our approach:

\textbf{(1) Number of end-users:} We randomly select different numbers of end-users $ n = 4, 8, 16, ..., 512$. For each setting, we run the experiment 100 times to get 100 different random end-user distributions so that extreme cases, such as overly dense or sparse user distributions, are properly neutralized.

\textbf{(2) Number of edge servers:} The $ n $ end-users are located within the combined coverage of $ M $ edge servers. We assume that a total of $ m $ servers, where $ m =  10\%, 20\%, ..., 100\% * M $, are available for accommodating those $ n $ end-users.

\textbf{(3) Remaining server capacity:} We experiment various levels of remaining server capacity based on the combined user workload. To be specific, we calculate 100\%, 150\%, ..., 300\% of the combined user workload, then normally distribute it to $ M $ edge servers collectively covering the $ n $ end-users. 

\textbf{Performance metrics:} We evaluate the three approaches, namely our VSVBP, the random and the greedy baseline approaches, using the following metrics: (1) the percentage of allocated end-users of all end-users, the higher the better; (2) the percentage of hired edge servers of all available edge servers, the lower the better; and (3) the execution time (CPU time), the lower the better.

Given the data and the experiment parameters, we conduct three sets of experiments. The corresponding settings are described in Table 2. For each set, we vary one parameter and keep the other two fixed to observe the impact of each parameter on the approaches in the evaluation metrics.

\begin{table}
	\caption{Experiment Settings}
	\begin{tabularx}{\textwidth}{ |c| *{3}{Y|} }
		\hline
		\textbf{Factor} & \textbf{Number of users} & \textbf{Percentage of the total number of servers} & \textbf{Remaining server capacity} \\
		\hline
		\textbf{Set \#1} & $ 4, 8,..., 512 $ & $ 100\% $ & $ 300\% $ \\
		\hline
		\textbf{Set \#2} & $ 512 $ & $ 10\%, 20\%,..., 100\% $ & $ 300\% $ \\
		\hline
		\textbf{Set \#3} & $ 512 $ & $ 100\% $ & $ 100\%, 150\%,..., 300\% $ \\
		\hline
	\end{tabularx}
\end{table}

In experiment set 1, the number of users vary from 4, 8, 16, 32, 64, 128, 256 to 512. All the edge servers, which have end-users within their coverage, can serve those end-users. The total remaining server capacity is 300\% of the combined user workload. In experiment set 2, the number of users is fixed at 512, and the total remaining server capacity is fixed at 300\% of the combined user workload. We change the number of edge servers that would be used to accommodate end-user, i.e., 10\%, 20\%, ..., 100\% of all edge servers to be made available for hire. In the last experiment set, we keep the number of users fixed at 512 and make all edge servers available for hire. The changing factor is the remaining server capacity -- 100\%, 150\%,..., 300\% of all users' workload combined.

\subsection{Experimental results and discussion}

Figure 2, 3 and 4 show the results of the experiment set 1, 2 and 3, respectively. The three performance metrics are depicted in each sub-figure: (a) percentage of user allocated, (b) percentage of servers hired, and (c) execution time.

Figure 2 shows that in experiment set 1, as we increase the number of end-users from 4 users to 512, the random approach performs poorly in terms of allocated users percentage (only 20\% - 25\% of the users are allocated) compared to the greedy approach and our approach, which give an equal performance with all users having been allocated. However, in terms of the number of edge servers hired, our approach starts to outperform the greedy approach as the number of end-users exceeds 32. The percentages of servers hired by the greedy and the random methods keep growing as the number of end-users increases, up to around 87.04\% when serving 512 end-users. By contrast, our approach stably uses only around 32\% of all available edge servers, 2.7 times less than that of the greedy approach, and remains steady even when the number of end-users increases from 32 to 512. 

\newlength\figureheight
\newlength\figurewidth
\setlength\figureheight{1.5in}
\setlength\figurewidth{2in}
\setlength{\belowcaptionskip}{-2ex}	
\begin{figure}
	\begin{subfigure}[b]{0.325\textwidth}
		\centering
		\resizebox{\linewidth}{!}{
\begin{tikzpicture}

\definecolor{color0}{rgb}{0.12156862745098,0.466666666666667,0.705882352941177}
\definecolor{color1}{rgb}{1,0.498039215686275,0.0549019607843137}
\definecolor{color2}{rgb}{0.172549019607843,0.627450980392157,0.172549019607843}

\begin{axis}[
xlabel={Number of users},
ylabel={Assigned users pct.},
xmin=-21.4, xmax=537.4,
ymin=13.703125, ymax=104.109375,
width=\figurewidth,
height=\figureheight,
ytick={0,20,40,60,80,100,120},
yticklabels={0\%, 20\%, 40\%, 60\%, 80\%,100\%,120\%},
xtick={32,128,256,512},
xticklabels={32,128,256,512},
tick align=outside,
tick pos=left,
xmajorgrids,
x grid style={white!69.019607843137251!black},
ymajorgrids,
y grid style={white!69.019607843137251!black},
legend style={at={(1,0.5)},anchor=east,font=\fontsize{4}{5}\selectfont},
]
\addplot [semithick, color0, mark=*, mark size=2, mark options={solid}]
table {%
4 77.5
8 67.5
16 79.375
32 93.4375
64 98.28125
128 99.375
256 99.9609375
512 100
};
\addlegendentry{Our method}

\addplot [semithick, color1, mark=square*, mark size=2, mark options={solid}]
table {%
4 35
8 25
16 20
32 17.8125
64 19.84375
128 22.65625
256 25.15625
512 24.140625
};
\addlegendentry{Random}

\addplot [semithick, color2, mark=x, mark size=2, mark options={solid}]
table {%
4 75
8 62.5
16 75
32 90.3125
64 97.1875
128 99.21875
256 99.9609375
512 100
};
\addlegendentry{Greedy}

\end{axis}
\end{tikzpicture}}
		\caption{Pct. of users allocated}
		\label{figS1:A}
	\end{subfigure}
	\begin{subfigure}[b]{0.325\textwidth}
		\centering
		\resizebox{\linewidth}{!}{
\begin{tikzpicture}

\definecolor{color0}{rgb}{0.12156862745098,0.466666666666667,0.705882352941177}
\definecolor{color1}{rgb}{1,0.498039215686275,0.0549019607843137}
\definecolor{color2}{rgb}{0.172549019607843,0.627450980392157,0.172549019607843}

\begin{axis}[
xlabel={Number of users},
ylabel={Hired servers pct.},
xmin=-21.4, xmax=537.4,
ymin=0.20331453362256, ymax=91.1750802603037,
width=\figurewidth,
height=\figureheight,
ytick={0,20,40,60,80,100},
yticklabels={  0\%, 20\%, 40\%, 60\%, 80\%,100\%},
xtick={32,128,256,512},
xticklabels={32,128,256,512},
tick align=outside,
tick pos=left,
xmajorgrids,
x grid style={white!69.019607843137251!black},
ymajorgrids,
y grid style={white!69.019607843137251!black}
]
\addplot [semithick, color0, mark=*, mark size=2, mark options={solid}, forget plot]
table {%
4 14.572864321608
8 11.7136659436009
16 18.3673469387755
32 27.8738555442523
64 32.603201347936
128 31.4262691377921
256 31.84
512 31.6
};
\addplot [semithick, color1, mark=square*, mark size=2, mark options={solid}, forget plot]
table {%
4 7.03517587939699
8 4.33839479392625
16 4.66472303206997
32 5.59511698880977
64 10.2780117944398
128 21.0314262691378
256 39.28
512 59.68
};
\addplot [semithick, color2, mark=x, mark size=2, mark options={solid}, forget plot]
table {%
4 14.070351758794
8 10.8459869848156
16 17.4927113702624
32 27.8738555442523
64 41.7017691659646
128 57.2925060435133
256 73.92
512 87.04
};
\end{axis}

\end{tikzpicture}}
		\caption{Pct. of servers hired}
		\label{figS1:B}
	\end{subfigure}
	\begin{subfigure}[b]{0.325\textwidth}
		\centering
		\resizebox{\linewidth}{!}{
\begin{tikzpicture}

\definecolor{color0}{rgb}{0.12156862745098,0.466666666666667,0.705882352941177}
\definecolor{color1}{rgb}{1,0.498039215686275,0.0549019607843137}
\definecolor{color2}{rgb}{0.172549019607843,0.627450980392157,0.172549019607843}

\begin{axis}[
xlabel={Number of users},
ylabel={CPU Time (s)},
xtick={32,128,256,512},
xticklabels={32,128,256,512},
xmin=-21.4, xmax=537.4,
ymin=-1.14575501323308, ymax=24.2834153067715,
width=\figurewidth,
height=\figureheight,
tick align=outside,
tick pos=left,
xmajorgrids,
x grid style={white!69.019607843137251!black},
ymajorgrids,
y grid style={white!69.019607843137251!black},
]
\addplot [semithick, color0, mark=*, mark size=2, mark options={solid}]
table {%
	4 0.0375635254146459
	8 0.0801195692197155
	16 0.196309797943229
	32 0.525388576497403
	64 1.29684488129469
	128 3.41969692121202
	256 8.42581357970812
	512 23.1275439285895
};

\addplot [semithick, color1, mark=square*, mark size=2, mark options={solid}]
table {%
	4 0.0101163649489465
	8 0.0190608030644853
	16 0.0382665449812066
	32 0.0792906401586606
	64 0.16116854762904
	128 0.301681324818765
	256 0.669727022584084
	512 1.50816555711863
};

\addplot [semithick, color2, mark=x, mark size=2, mark options={solid}]
table {%
	4 0.0142761569401472
	8 0.0352728771492821
	16 0.0524709364961268
	32 0.100033962710131
	64 0.19626653388782
	128 0.397835773229599
	256 0.767182037993007
	512 1.41130680913921
};

\end{axis}
\end{tikzpicture}}
		\caption{Execution time}
		\label{figS1:C}
	\end{subfigure}
	\caption{Resulted metrics of set \#1 (Number of users changing)}
\end{figure}
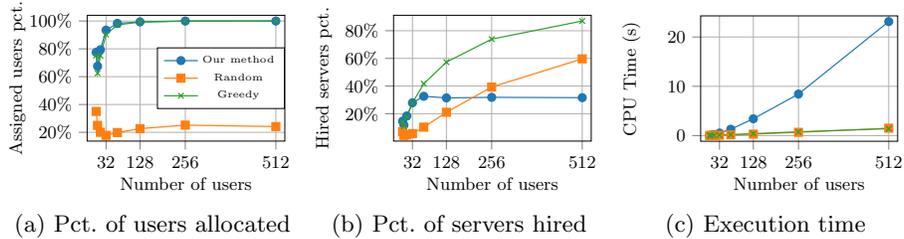
\vspace{0.00mm}
\begin{figure}
	\begin{subfigure}[b]{0.325\textwidth}
		\centering
		\resizebox{\linewidth}{!}{
\begin{tikzpicture}

\definecolor{color0}{rgb}{0.12156862745098,0.466666666666667,0.705882352941177}
\definecolor{color1}{rgb}{1,0.498039215686275,0.0549019607843137}
\definecolor{color2}{rgb}{0.172549019607843,0.627450980392157,0.172549019607843}

\begin{axis}[
xlabel={Percentage of total no. of servers},
ylabel={Assigned users pct.},
xmin=5.5, xmax=104.5,
ymin=1.15234375, ymax=104.70703125,
width=\figurewidth,
height=\figureheight,
ytick={0,20,40,60,80,100,120},
yticklabels={  0\%, 20\%, 40\%, 60\%, 80\%,100\%,120\%},
xticklabels={,,20\%,40\%,60\%,80\%,100\%},
tick align=outside,
tick pos=left,
xmajorgrids,
x grid style={white!69.019607843137251!black},
ymajorgrids,
y grid style={white!69.019607843137251!black}
]
\addplot [semithick, color0, mark=*, mark size=2, mark options={solid}, forget plot]
table {%
10 26.07421875
20 51.484375
30 74.27734375
40 87.0703125
50 92.59765625
60 97.0703125
70 98.7890625
80 99.453125
90 99.78515625
100 100
};
\addplot [semithick, color1, mark=square*, mark size=2, mark options={solid}, forget plot]
table {%
10 5.859375
20 9.21875
30 11.34765625
40 13.28125
50 16.25
60 18.0859375
70 18.7890625
80 21.0546875
90 24.00390625
100 24.140625
};
\addplot [semithick, color2, mark=x, mark size=2, mark options={solid}, forget plot]
table {%
10 26.03515625
20 50.83984375
30 72.67578125
40 85.60546875
50 91.796875
60 96.50390625
70 98.76953125
80 99.4140625
90 99.74609375
100 100
};
\end{axis}

\end{tikzpicture}}
		\caption{Pct. of users allocated}
		\label{figS2:A}
	\end{subfigure}
	\begin{subfigure}[b]{0.325\textwidth}
		\centering
		\resizebox{\linewidth}{!}{
\begin{tikzpicture}

\definecolor{color0}{rgb}{0.12156862745098,0.466666666666667,0.705882352941177}
\definecolor{color1}{rgb}{1,0.498039215686275,0.0549019607843137}
\definecolor{color2}{rgb}{0.172549019607843,0.627450980392157,0.172549019607843}

\begin{axis}[
xlabel={Percentage of total no. of servers},
ylabel={Hired servers pct.},
xmin=5.5, xmax=104.5,
ymin=28.18, ymax=103.42,
width=\figurewidth,
height=\figureheight,
ytick={20,30,40,50,60,70,80,90,100,110},
yticklabels={ 20\%, 30\%, 40\%, 50\%, 60\%, 70\%, 80\%, 90\%,100\%,110\%},
xticklabels={,,20\%,40\%,60\%,80\%,100\%},
tick align=outside,
tick pos=left,
xmajorgrids,
x grid style={white!69.019607843137251!black},
ymajorgrids,
y grid style={white!69.019607843137251!black}
]
\addplot [semithick, color0, mark=*, mark size=2, mark options={solid}, forget plot]
table {%
10 100
20 98.8
30 92.8947368421052
40 82.8
50 67.258064516129
60 56.2666666666667
70 47.7272727272727
80 41.6
90 36.3392857142857
100 31.6
};
\addplot [semithick, color1, mark=square*, mark size=2, mark options={solid}, forget plot]
table {%
10 84.1666666666667
20 76.4
30 72.8947368421053
40 66.8
50 68.0645161290323
60 64
70 61.7045454545455
80 62
90 63.0357142857143
100 59.68
};
\addplot [semithick, color2, mark=x, mark size=2, mark options={solid}, forget plot]
table {%
10 100
20 100
30 100
40 99.4
50 98.5483870967742
60 97.8666666666667
70 96.3636363636364
80 93.5
90 90.2678571428571
100 87.04
};
\end{axis}

\end{tikzpicture}}
		\caption{Pct. of servers hired}
		\label{figS2:B}
	\end{subfigure}
	\begin{subfigure}[b]{0.325\textwidth}
		\centering
		\resizebox{\linewidth}{!}{
\begin{tikzpicture}

\definecolor{color0}{rgb}{0.12156862745098,0.466666666666667,0.705882352941177}
\definecolor{color1}{rgb}{1,0.498039215686275,0.0549019607843137}
\definecolor{color2}{rgb}{0.172549019607843,0.627450980392157,0.172549019607843}

\begin{axis}[
xlabel={Percentage of total no. of servers},
ylabel={CPU Time (s)},
xticklabels={,,20\%,40\%,60\%,80\%,100\%},
xmin=5.5, xmax=104.5,
ymin=-0.837266343167812, ymax=24.2687253701018,
width=\figurewidth,
height=\figureheight,
tick align=outside,
tick pos=left,
xmajorgrids,
x grid style={white!69.019607843137251!black},
ymajorgrids,
y grid style={white!69.019607843137251!black}
]
\addplot [semithick, color0, mark=*, mark size=2, mark options={solid}, forget plot]
table {%
10 1.01938076080733
20 2.06705356849488
30 3.28728151972064
40 5.07666625279926
50 7.56957227038547
60 10.7701201537537
70 13.543216737984
80 16.6778257636171
90 17.2548353128557
100 23.1275439285895
};
\addplot [semithick, color1, mark=square*, mark size=2, mark options={solid}, forget plot]
table {%
10 0.303915098344442
20 0.436582873493353
30 0.6002793390242
40 0.708399882751473
50 0.899013545403614
60 0.973269597785111
70 1.06404343130889
80 1.10841955211108
90 1.28560783224493
100 1.50816555711863
};
\addplot [semithick, color2, mark=x, mark size=2, mark options={solid}, forget plot]
table {%
10 0.8592104011228
20 1.10279795558126
30 1.25663938983162
40 1.35803958362394
50 1.35114801213567
60 1.37812285199725
70 1.40573052466807
80 1.41322497692745
90 1.41596767910523
100 1.41130680913921
};
\end{axis}

\end{tikzpicture}}
		\caption{Execution time}
		\label{figS2:C}
	\end{subfigure}
	\caption{Resulted metrics of set \#2 (Number of servers changing)}
\end{figure}
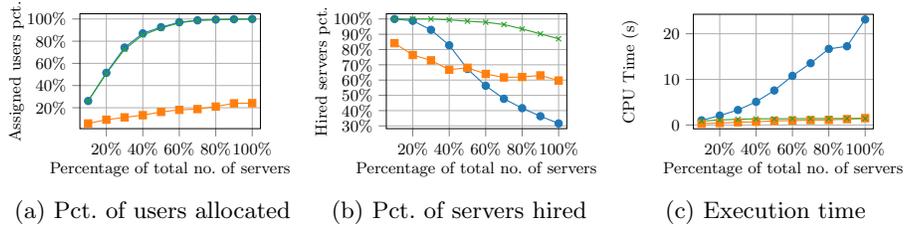
\vspace{0.00mm}
\begin{figure}
	\begin{subfigure}[b]{0.325\textwidth}
		\centering
		\resizebox{\linewidth}{!}{
\begin{tikzpicture}

\definecolor{color0}{rgb}{0.12156862745098,0.466666666666667,0.705882352941177}
\definecolor{color1}{rgb}{1,0.498039215686275,0.0549019607843137}
\definecolor{color2}{rgb}{0.172549019607843,0.627450980392157,0.172549019607843}

\begin{axis}[
xlabel={Combined user workload percentage},
xticklabels={,,100\%,150\%,200\%,250\%,300\%},
ylabel={Assigned users pct.},
xmin=90, xmax=310,
ymin=4.6796875, ymax=104.5390625,
width=\figurewidth,
height=\figureheight,
ytick={0,20,40,60,80,100,120},
yticklabels={  0\%, 20\%, 40\%, 60\%, 80\%,100\%,120\%},
tick align=outside,
tick pos=left,
xmajorgrids,
x grid style={white!69.019607843137251!black},
ymajorgrids,
y grid style={white!69.019607843137251!black}
]
\addplot [semithick, color0, mark=*, mark size=2, mark options={solid}, forget plot]
table {%
100 83.49609375
150 99.84375
200 100
250 100
300 100
};
\addplot [semithick, color1, mark=square*, mark size=2, mark options={solid}, forget plot]
table {%
100 9.21875
150 14.27734375
200 17.87109375
250 21.953125
300 24.140625
};
\addplot [semithick, color2, mark=x, mark size=2, mark options={solid}, forget plot]
table {%
100 81.11328125
150 99.16015625
200 99.9609375
250 100
300 100
};
\end{axis}

\end{tikzpicture}}
		\caption{Pct. of users allocated}
		\label{figS3:A}
	\end{subfigure}
	\begin{subfigure}[b]{0.325\textwidth}
		\centering
		\resizebox{\linewidth}{!}{
\begin{tikzpicture}

\definecolor{color0}{rgb}{0.12156862745098,0.466666666666667,0.705882352941177}
\definecolor{color1}{rgb}{1,0.498039215686275,0.0549019607843137}
\definecolor{color2}{rgb}{0.172549019607843,0.627450980392157,0.172549019607843}

\begin{axis}[
xlabel={Combined user workload percentage},
xticklabels={,,100\%,150\%,200\%,250\%,300\%},
ylabel={Hired servers pct.},
xmin=90, xmax=310,
ymin=25.984, ymax=100.256,
width=\figurewidth,
height=\figureheight,
ytick={20,30,40,50,60,70,80,90,100,110},
yticklabels={ 20\%, 30\%, 40\%, 50\%, 60\%, 70\%, 80\%, 90\%,100\%,110\%},
tick align=outside,
tick pos=left,
xmajorgrids,
x grid style={white!69.019607843137251!black},
ymajorgrids,
y grid style={white!69.019607843137251!black}
]
\addplot [semithick, color0, mark=*, mark size=2, mark options={solid}, forget plot]
table {%
100 96.4
150 65.76
200 47.36
250 38.08
300 31.6
};
\addplot [semithick, color1, mark=square*, mark size=2, mark options={solid}, forget plot]
table {%
100 29.36
150 41.6
200 47.2
250 55.12
300 59.68
};
\addplot [semithick, color2, mark=x, mark size=2, mark options={solid}, forget plot]
table {%
100 96.88
150 95.6
200 92.32
250 89.12
300 87.04
};
\end{axis}

\end{tikzpicture}}
		\caption{Pct. of servers hired}
		\label{figS3:B}
	\end{subfigure}
	\begin{subfigure}[b]{0.325\textwidth}
		\centering
		\resizebox{\linewidth}{!}{
\begin{tikzpicture}

\definecolor{color0}{rgb}{0.12156862745098,0.466666666666667,0.705882352941177}
\definecolor{color1}{rgb}{1,0.498039215686275,0.0549019607843137}
\definecolor{color2}{rgb}{0.172549019607843,0.627450980392157,0.172549019607843}

\begin{axis}[
xlabel={Combined user workload percentage},
xticklabels={,,100\%,150\%,200\%,250\%,300\%},
ylabel={CPU Time (s)},
xmin=90, xmax=310,
ymin=0.111491575242744, ymax=24.223546421606,
width=\figurewidth,
height=\figureheight,
tick align=outside,
tick pos=left,
xmajorgrids,
x grid style={white!69.019607843137251!black},
ymajorgrids,
y grid style={white!69.019607843137251!black}
]
\addplot [semithick, color0, mark=*, mark size=2, mark options={solid}, forget plot]
table {%
100 10.8135104599598
150 15.3444166115205
200 18.5641499274447
250 19.8473232819073
300 23.1275439285895
};
\addplot [semithick, color1, mark=square*, mark size=2, mark options={solid}, forget plot]
table {%
100 1.20749406825926
150 1.51206196743879
200 1.41280569105766
250 1.45010033079452
300 1.50816555711863
};
\addplot [semithick, color2, mark=x, mark size=2, mark options={solid}, forget plot]
table {%
100 1.48058480714863
150 1.42834917762011
200 1.45005565873907
250 1.41689440029259
300 1.41130680913921
};
\end{axis}

\end{tikzpicture}}
		\caption{Execution time}
		\label{figS3:C}
	\end{subfigure}
	\caption{Resulted metrics of set \#3 (Remaining server capacity changing)}
\end{figure}
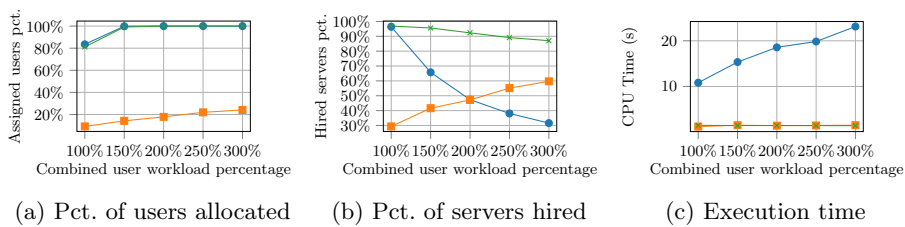

In experiment set 2, we change the number of edge servers available for hire. As depicted in Fig. 3(a), the allocated user percentage follows a similar trend as in experiment set 1. Regarding the percentage of hired edge servers (Fig. 3(b)), our approach continues to outperform the other two approaches as the number of edge servers increases.

Figure 4 shows that, in the last set of experiments, where the edge servers' total remaining capacity increases, we can observe the same trending patterns with our approach being the most effective out of the three approaches studied. As we increase the combined user workload percentage from 100\% to 300\%, our approach uses significantly fewer servers, dropping from 96.4\% to 31.6\% while the greedy method has to use around 90\% of the all available servers.

Note that in all three set of experiments, the random approach seems to perform better than ours with fewer hired servers on some occasions. For example, when the number of users varies between 4 and 128 (Fig.2(b)), when the total number of hired edge servers percentage varies between 10\% and 50\% (Fig. 3(b)), and when the total remaining server capacity changes between 100\% and 200\% (Fig. 4(b)). In fact, the random approach does not produce better results in these scenarios because although it uses fewer servers, the number of users allocated is extremely small (only 6\% - 20\% of all end-users experimenting) compared to the other approaches, as shown in Fig. 2, 3, 4(a).

In terms of efficiency, the computation time of our approach increases considerably as we increase any one of the three parameters. In experiment set 1 with 512 users, the greedy and random methods take only approximately 1.5 seconds while our approach takes around 23.1 seconds to solve an instance of the EUA problem. This can also be observed in experiment sets 2 and 3, where we increase the number of servers available for hired and the total remaining server capacity respectively. Since the EUA problem is an $\mathcal{NP}$-hard problem, it is expected that our approach, which optimally solves the problem, will take the most time as opposed to the other approaches, which can only make local decisions without considering the problem globally. 

In general, increasing one of the three experimental parameters will increase the complexity of the EUA problem, which is an $\mathcal{NP}$-hard problem, and thus take more time to find an optimal solution. Our experimental results show that the random approach is not able to maximize the number of users allocated (the first optimization objective) as it can assign around only 20\% of all the end-users in the experiments. The greedy approach is able to assign a similar number of end-users as our approach; the edge servers' adequate remaining capacities allow the greedy approach to find a capable edge server to accommodate most end-users. However, as shown in Fig. 2, 3, 4(b), our approach hires much fewer edge servers (the second optimization objective) than the greedy method to accommodate all the end-users. This is shown in all three experiment results, especially as the EUA problem scales up.

\subsection{Threats to Validity}
\textit{Threats to construct validity.} The main threat to the construct validity in our study lies in the comparison with the two baseline methods, i.e., the random and greedy methods. The EUA problem studied in this research is a problem that has not been investigated before in this domain. Thus, we selected these two common and intuitive methods as baselines in our evaluation. Their designs are simple, especially the random method, which employs no heuristics. As a result, our approach is likely to obtain better experimental results, leading to a threat where the comparison with the selected baselines might not properly demonstrate the effectiveness of our approach in solving the EUA problem. To minimize this threat, we conducted experiments with three changing parameters as described in Table 2 to simulate different service deployment scenarios in the real world. This way, we could reliably evaluate our approach through both comparison with the baseline methods and also impacts of varying each experimental parameters on our approach.

\textit{Threats to external validity.} A major threat to external validity is whether our findings based on the experimental dataset can be generalized to other application domains in edge computing. Since there is currently no real-world dataset for this type of edge computing problems, we synthesized a dataset of edge servers and end-users based on reliable real-world data sources (ACMA and APNIC). However, this is a generic dataset, and it is possible that different application domains might have different factors that could impact the experimental results, such as the density and distributions of edge servers and end-users. Thus, our approach was evaluated across a breadth of problem scoping, varying in size, i.e., number of end-users, and complexity, i.e., number of edge servers and edge servers' remaining capacities, to simulate as many types of edge server and end-user density and distribution as possible, as well as their combinations. This helped reduce the threat to the external validity of our evaluation and increased the generalizability of our results.

\textit{Threats to internal validity.} A threat to internal validity of our work is the comprehensiveness of our experiments and whether or not the results are not biased by the experimental parameter settings. To mitigate this threat, we carried out extensive experiments with systematically selected parameters. The three experimental parameters (discussed in section 5.2) are the three representative parameters that directly impact the outcomes of the approaches. Also, for each experiment set, we experimented with 100 different user distributions randomly selected from the pool of users to eliminate the potential bias caused by highly special scenarios such as overly dense or sparse distributions. Another threat to the internal validity of our evaluation is where more sophisticated scenarios could be simulated, e.g., those where two or more of those parameters change at the same time. In those scenarios, the results can be predicted in general based on the results that we have obtained. For example, if the total number of available edge servers and their total remaining capacities increase at the same time, the percentage of used servers of all hired by our approach will decline with a trend similar to but more significant than those shown in Fig. 3(b) and Fig. 4(b).

\textit{Threats to conclusion validity.} The lack of statistical tests is the biggest threat to our conclusion validity. Statistical tests will be included in our future work to prove a statistically significant relationship between the experiment settings and the results. In this paper, we have compensated for this with meaningful comparison baselines and extensive experiments that cover many different scenarios, varying in both size and complexity. When an experimental parameter changes, the results are averaged over 100 runs of the experiment.
\section{Related Work}

Resource management in cloud computing has been extensively investigated in the last decade in many research tracks such as load balancing \cite{Mitzenmacher2001}, virtual machine placement and provisioning \cite{Wolke2015}, server and task allocation \cite{Ren2017}, etc. 

Edge computing, or fog computing, is a new computing paradigm coined by Cisco in 2012 \cite{Bonomi2012}. Edge computing is a natural extension of cloud computing with regard to the network topology and infrastructure deployment, where the architecture is more geographically distributed compared to cloud computing. This new architecture pushes the cloud resources closer to end-users. Barcelona, Spain is one of the first cities implementing edge computing with many applications, including power monitoring in public spaces, access control and telemetry of sensors, event-based video streaming, traffic analysis and regulation, and connectivity on demand \cite{Yannuzzi2017}. There are more than 3,000 edge servers deployed across the city serving thousands of IoT devices. The sheer number of edge servers and end-devices, with the horizontal scaling nature of edge computing, leads to the need for effective and efficient resource allocation solutions.

Chen et al. \cite{Chen2016} proposed a distributed game theoretic computation offloading algorithm that was able to achieve a Nash equilibrium, minimizing the total energy consumption and offloading latency in the multi-channel mobile edge computing environment. By the proposed approach, they were able to optimally decide whether the users should offload computing tasks to an edge server and if yes, which wireless channel to be used for the computation offloading. In \cite{Yao2017}, Yao et al. tackled the problem of cost-effective edge server deployment using integer linear programming. They took into account the factors of resource capacity, user-server latency, and deployment costs. In their research, each edge server might not have the service installed to fulfill the requests from end-users. Thus, users' requests might have to travel across different edge servers until executed. However, they assumed that each server covers a region exclusively with other servers. They also assumed a predetermined edge server that first receives the user request. Our research targets more realistic edge computing scenarios where different edge servers' coverages might partially overlap. The authors of \cite{Wang2017a} also made an assumption that each small geographical area will only receive coverage from only a single edge server, which will be unlikely to happen in real-world scenarios. In \cite{Tran2017}, the authors formulated a problem similar to the EUA problem but with different objectives, which are to reduce task completion time and energy consumption. Yin et al. \cite{Yin2017a} addressed the edge server placement and provisioning problem with the objective of maximizing users coverage and minimizing network latency.

To the best of our knowledge, our work is the first to tackle the EUA problem in scenarios with multiple edge servers and end-users that possess and require multi-dimensional computing capacities. We also realistically and innovatively address this problem with respect to proximity constraints with the aims to maximize the number of allocated users and minimize the number of hired servers.
\section{Conclusion}

Edge computing is a promising new computing architecture, especially for high volume, data processing-intensive, latency-sensitive applications and services. However, when an edge computing scenario scales up, an ineffective edge user allocation solution will greatly increase the operational costs for service providers. To address this problem, we formulated the edge user allocation (EUA) problem as a variant of the bin packing problem named variable sized vector bin packing, an $\mathcal{NP}$-hard problem. We solved this problem using a Lexicographic Goal Programming technique with two optimization objectives, i.e., to maximize the number of users allocated and minimize the number of edge servers hired. We then conducted extensive experiments in scenarios with various service deployment requirements. Our experimental results show that our approach significantly outperforms two baseline approaches, greedy and random. It is capable of allocating the most end-users with significantly fewer edge servers -- nearly three times less than the greedy method -- as the EUA problem scales up.

This research has established a basic foundation for the EUA problem and opened up a number of research directions. In our future work, we will take into account the users' mobility as well as the dynamics of users' computation tasks. In addition, apart from the proximity and capacity constraints, there are several elements that also play an important role such as network latency, service availability, pricing, and security. \newline

\textbf{Acknowledgments.} This research is funded by Australian Research Council Discovery Projects (DP170101932 and DP18010021).

\bibliographystyle{splncs04}
\bibliography{IEEEabrv,references}

\begin{thebibliography}{10}
\providecommand{\url}[1]{\texttt{#1}}
\providecommand{\urlprefix}{URL }
\providecommand{\doi}[1]{https://doi.org/#1}

\bibitem{Bonomi2012}
Bonomi, F., Milito, R., Zhu, J., Addepalli, S.: {Fog Computing and Its Role in
  the Internet of Things}. In: Proceedings of the First Edition of the MCC
  Workshop on Mobile Cloud Computing. pp. 13--16. MCC '12, ACM, New York, NY,
  USA (2012). \doi{10.1145/2342509.2342513}

\bibitem{Chen2016}
Chen, X., Jiao, L., Li, W., Fu, X.: {Efficient Multi-User Computation
  Offloading for Mobile-Edge Cloud Computing}. IEEE/ACM Transactions on
  Networking  \textbf{24}(5),  2795--2808 (2016).
  \doi{10.1109/TNET.2015.2487344}

\bibitem{Garey1979}
Garey, M.R., Johnson, D.S.: {Computers and intractability}, vol.~29. W. H.
  Freeman and Company, New York (2002)

\bibitem{Ericsson2017}
Heuveldop, N.: {Ericsson Mobility Report}. Tech. Rep. November, Ericsson
  (2017),
  \url{https://www.ericsson.com/assets/local/mobility-report/documents/2017/ericsson-mobility-report-november-2017.pdf}

\bibitem{Hu2015}
Hu, Y.C., Patel, M., Sabella, D., Sprecher, N., Young, V.: {Mobile Edge
  Computing A key technology towards 5G}. Tech. Rep.~11, European
  Telecommunications Standards Institute (2015),
  \url{http://www.etsi.org/images/files/ETSIWhitePapers/
  etsi{\_}wp11{\_}mec{\_}a{\_}key{\_}technology{\_}towards{\_}5g.pdf}

\bibitem{Lin2017a}
Lin, Y., Shen, H.: {CloudFog: Leveraging Fog to Extend Cloud Gaming for
  Thin-Client MMOG with High Quality of Service}. IEEE Transactions on Parallel
  and Distributed Systems  \textbf{28}(2),  431--445 (2017).
  \doi{10.1109/TPDS.2016.2563428}

\bibitem{Mitzenmacher2001}
Mitzenmacher, M.: {The power of two choices in randomized load balancing}. IEEE
  Transactions on Parallel and Distributed Systems  \textbf{12}(10),
  1094--1104 (oct 2001). \doi{10.1109/71.963420}

\bibitem{Ren2017}
Ren, R., Tang, X., Li, Y., Cai, W.: {Competitiveness of dynamic bin packing for
  online cloud server allocation}. IEEE/ACM Transactions on Networking
  \textbf{25}(3),  1324--1331 (jun 2017). \doi{10.1109/TNET.2016.2630052}

\bibitem{romero2014}
Romero, C.: {Handbook of critical issues in goal programming}. Elsevier (2014)

\bibitem{BI2017}
Smith, J.: {The Mobile Gaming Report}. Tech. rep., Business Insider
  Intelligence (2016),
  \url{http://www.businessinsider.com/the-mobile-gaming-report-market-size-the-free-to-play-model-and-new-opportunities-to-market-and-monetize}

\bibitem{Tran2017}
Tran, T.X., Pompili, D.: {Joint Task Offloading and Resource Allocation for
  Multi-Server Mobile-Edge Computing Networks}. CoRR  \textbf{abs/1705.0}
  (2017), \url{http://arxiv.org/abs/1705.00704}

\bibitem{Varghese2017}
Varghese, B., Wang, N., Nikolopoulos, D.S., Buyya, R.: {Feasibility of Fog
  Computing}. CoRR  \textbf{abs/1701.0} (2017),
  \url{http://arxiv.org/abs/1701.05451}

\bibitem{Wang2017a}
Wang, L., Jiao, L., Li, J., M{\"{u}}hlh{\"{a}}user, M.: {Online Resource
  Allocation for Arbitrary User Mobility in Distributed Edge Clouds}. In: 2017
  IEEE 37th International Conference on Distributed Computing Systems (ICDCS).
  pp. 1281--1290 (jun 2017). \doi{10.1109/ICDCS.2017.30}

\bibitem{Wolke2015}
Wolke, A., Tsend-Ayush, B., Pfeiffer, C., Bichler, M.: {More than bin packing:
  Dynamic resource allocation strategies in cloud data centers}. Information
  Systems  \textbf{52},  83--95 (2015). \doi{10.1016/j.is.2015.03.003}

\bibitem{Yannuzzi2017}
Yannuzzi, M., {Van Lingen}, F., Jain, A., Parellada, O.L., Flores, M.M.,
  Carrera, D., Perez, J.L., Montero, D., Chacin, P., Corsaro, A., Olive, A.: {A
  new era for cities with fog computing}. IEEE Internet Computing
  \textbf{21}(2),  54--67 (2017). \doi{10.1109/MIC.2017.25}

\bibitem{Yao2017}
Yao, H., Bai, C., Xiong, M., Zeng, D., Fu, Z.: {Heterogeneous cloudlet
  deployment and user-cloudlet association toward cost effective fog
  computing}. Concurrency Computation  \textbf{29}(16), ~1--9 (2017).
  \doi{10.1002/cpe.3975}

\bibitem{Yi2015}
Yi, S., Li, C., Li, Q.: {A Survey of Fog Computing: Concepts, Applications and
  Issues}. In: Proceedings of the 2015 Workshop on Mobile Big Data - Mobidata
  '15. pp. 37--42. Mobidata '15, ACM, New York, NY, USA (2015).
  \doi{10.1145/2757384.2757397}

\bibitem{Yin2017a}
Yin, H., Zhang, X., Liu, H.H., Luo, Y., Tian, C., Zhao, S., Li, F.: {Edge
  Provisioning with Flexible Server Placement}. IEEE Transactions on Parallel
  and Distributed Systems  \textbf{28}(4),  1031--1045 (apr 2017).
  \doi{10.1109/TPDS.2016.2604803}

\end{thebibliography}

\end{document}